\documentstyle[12pt]{article}

\catcode`\@=11
\long\def\@makefntext#1{
\protect\noindent \hbox to 3.2pt {\hskip-.9pt
$^{{\ninerm\@thefnmark}}$\hfil}#1\hfill}		

\def\@makefnmark{\hbox to 0pt{$^{\@thefnmark}$\hss}}  

\def\ps@myheadings{\let\@mkboth\@gobbletwo
\def\@oddhead{\hbox{}
\rightmark\hfil\ninerm\thepage}
\def\@oddfoot{}\def\@evenhead{\ninerm\thepage\hfil
\leftmark\hbox{}}\def\@evenfoot{}
\def\sectionmark##1{}\def\subsectionmark##1{}}

\renewcommand{\thefootnote}{\fnsymbol{footnote}}

\newcounter{sectionc}\newcounter{subsectionc}\newcounter{subsubsectionc}
\renewcommand{\section}[1] {\vspace*{0.6cm}\addtocounter{sectionc}{1}
\setcounter{subsectionc}{0}\setcounter{subsubsectionc}{0}\noindent
	{\normalsize\bf\thesectionc. #1}\par\vspace*{0.4cm}}
\renewcommand{\subsection}[1] {\vspace*{0.6cm}\addtocounter{subsectionc}{1}
	\setcounter{subsubsectionc}{0}\noindent
	{\normalsize\it\thesectionc.\thesubsectionc. #1}\par\vspace*{0.4cm}}
\renewcommand{\subsubsection}[1]
{\vspace*{0.6cm}\addtocounter{subsubsectionc}{1}
	\noindent {\normalsize\rm\thesectionc.\thesubsectionc.\thesubsubsectionc.
	#1}\par\vspace*{0.4cm}}
\newcommand{\nonumsection}[1] {\vspace*{0.6cm}\noindent{\normalsize\bf #1}
	\par\vspace*{0.4cm}}

\newcounter{appendixc}
\newcounter{subappendixc}[appendixc]
\newcounter{subsubappendixc}[subappendixc]

\renewcommand{\appendix}[1] {\vspace*{0.6cm}
        \refstepcounter{appendixc}
        \setcounter{figure}{0}
        \setcounter{table}{0}
        \setcounter{equation}{0}
        \renewcommand{\thefigure}{\Alph{appendixc}.\arabic{figure}}
        \renewcommand{\thetable}{\Alph{appendixc}.\arabic{table}}
        \renewcommand{\theappendixc}{\Alph{appendixc}}
        \renewcommand{\theequation}{\Alph{appendixc}.\arabic{equation}}
        \noindent{\bf Appendix \theappendixc #1}\par\vspace*{0.4cm}}

\def\abstracts#1{{

\centering{\begin{minipage}{12.2truecm}\footnotesize\baselineskip=12pt\noindent
	\centerline{\footnotesize ABSTRACT}\vspace*{0.3cm}
	\parindent=0pt #1
	\end{minipage}}\par}}

\renewenvironment{thebibliography}[1]
	{\begin{list}{\arabic{enumi}.}
	{\usecounter{enumi}\setlength{\parsep}{0pt}
\setlength{\leftmargin 1.25cm}{\rightmargin 0pt}
	 \setlength{\itemsep}{0pt} \settowidth
	{\labelwidth}{#1.}\sloppy}}{\end{list}}

\newcounter{itemlistc}
\newcounter{romanlistc}
\newcounter{alphlistc}
\newcounter{arabiclistc}

\newcommand{\fcaption}[1]{
        \refstepcounter{figure}
        \setbox\@tempboxa = \hbox{\footnotesize Fig.~\thefigure. #1}
        \ifdim \wd\@tempboxa > 6in
           {\begin{center}
        \parbox{6in}{\footnotesize\baselineskip=12pt Fig.~\thefigure. #1}
            \end{center}}
        \else
             {\begin{center}
             {\footnotesize Fig.~\thefigure. #1}
              \end{center}}
        \fi}

\newcommand{\tcaption}[1]{
        \refstepcounter{table}
        \setbox\@tempboxa = \hbox{\footnotesize Table~\thetable. #1}
        \ifdim \wd\@tempboxa > 6in
           {\begin{center}
        \parbox{6in}{\footnotesize\baselineskip=12pt Table~\thetable. #1}
            \end{center}}
        \else
             {\begin{center}
             {\footnotesize Table~\thetable. #1}
              \end{center}}
        \fi}

\def\@citex[#1]#2{\if@filesw\immediate\write\@auxout
	{\string\citation{#2}}\fi
\def\@citea{}\@cite{\@for\@citeb:=#2\do
	{\@citea\def\@citea{,}\@ifundefined
	{b@\@citeb}{{\bf ?}\@warning
	{Citation `\@citeb' on page \thepage \space undefined}}
	{\csname b@\@citeb\endcsname}}}{#1}}

\newif\if@cghi
\def\cite{\@cghitrue\@ifnextchar [{\@tempswatrue
	\@citex}{\@tempswafalse\@citex[]}}
\def\citelow{\@cghifalse\@ifnextchar [{\@tempswatrue
	\@citex}{\@tempswafalse\@citex[]}}
\def\@cite#1#2{{$\null^{#1}$\if@tempswa\typeout
	{IJCGA warning: optional citation argument
	ignored: `#2'} \fi}}

 1
 1
 1

\font\ninerm=cmr9

\input{psfig}

\begin{document}
\hfill{\bf LTH 350}
\bigskip

\def\ga{\gamma}
\def\de{\delta}
\def\ep{\epsilon}
\def \la{\lambda}
\def \La{\Lambda}
\def\th{\theta}
\def\tautilde{\tilde\tau}
\def\chitilde{\tilde\chi}
\def\nutilde{\tilde\nu}
\def\gatilde{\tilde\ga}
\def\btilde{\tilde b}
\def\ttilde{\tilde t}
\def\tautilde{\tilde\tau}
\def\chitilde{\tilde\chi}
\def\nutilde{\tilde\nu}
\def\gatilde{\tilde\ga}
\def\btilde{\tilde b}
\def\dtilde{\tilde d}
\def\etilde{\tilde e}
\def\ttilde{\tilde t}
\def\utilde{\tilde u}

\def\bbar{{\overline{b}}}
\def\tbar{{\overline{t}}}
\def\taub{{\overline{\tau}}}
\def\half{{\textstyle{1\over2}}}
\def\sy{supersymmetry}
\def\sic{supersymmetric}
\def\sa{supergravity}
\def\ssm{supersymmetric standard model}
\def\sm{standard model}
\def\pa{\partial}
\def\semi{;\hfil\break}
\def\us#1{\bf{#1}}
\def\npb#1({{\it Nucl.\ Phys.}\ $\us {B#1}$\ (}
\def\zpc#1({{\it Zeit.\ f\"ur Physik}\ $\us {C#1}$\ (}
\def\prd#1({{\it Phys.\ Rev.}\ $\us  {D#1}$\ (}
\def\plb#1({{\it Phys.\ Lett.}\ $\us  {#1B}$\ (}

\newcommand{\st}{\scriptstyle}
\newcommand{\sst}{\scriptscriptstyle}
\newcommand{\mco}{\multicolumn}
\newcommand{\epp}{\epsilon^{\prime}}
\newcommand{\vep}{\varepsilon}
\newcommand{\ra}{\rightarrow}
\newcommand{\ppg}{\pi^+\pi^-\gamma}
\newcommand{\vp}{{\bf p}}
\newcommand{\ko}{K^0}
\newcommand{\kb}{\bar{K^0}}
\newcommand{\al}{\alpha}
\newcommand{\ab}{\bar{\alpha}}
\def\be{\begin{equation}}
\def\ee{\end{equation}}
\def\bea{\begin{eqnarray}}
\def\eea{\end{eqnarray}}
\def\CPbar{\hbox{{\rm CP}\hskip-1.80em{/}}}
\def\lf{16\pi^2}
\def\tr{{\rm tr}}
\def\Tr{{\rm Tr}}


\centerline{\normalsize\bf DRED, UNIVERSALITY AND THE
SUPERPARTICLE SPECTRUM
\footnote{Talk at PASCOS meeting, Baltimore, March 1995.}}

\vspace*{0.6cm}
\centerline{\footnotesize D.R.~TIMOTHY JONES}
\baselineskip=13pt
\centerline{\footnotesize\it DAMTP, University of Liverpool,
}
\baselineskip=12pt
\centerline{\footnotesize\it Liverpool, L69 3BX, UK}
\centerline{\footnotesize E-mail: drtj@amtp.liv.ac.uk}

\vspace*{0.5cm}
\abstracts{Recent work on the use of dimensional reduction
for the  regularisation of
 non--supersymmetric theories is reviewed. It is then shown that
there exists a class of theories for which a universal
form of the soft supersymmetry breaking terms is invariant under
renormalisation. It is argued that this universal form might be
approached as an
infra--red fixed point for the unified theory above the
unification scale.
The superparticle spectrum is calculated for these
theories.}

\normalsize\baselineskip=15pt
\setcounter{footnote}{0}
\renewcommand{\thefootnote}{\alph{footnote}}
\section{Introduction}

It is commonly assumed that the soft \sy\ terms  in the \ssm\ (SSM)
unify at high energies, and are determined ultimately by  four
parameters: $m_0, M, A$ and $B$ which we will define presently.  The
calculation of the sparticle spectrum in terms of  these parameters is a
major industry. At its most basic level, this  consists of
integrating the set of coupled differential equations  for the various
running masses and couplings from the scale of gauge  unification
($M_G$) down to $M_Z$, using the one--loop $\beta$--functions.  If we
wish to refine these calculations by including threshold  corrections or
using the two--loop $\beta$--functions then interesting issues  arise,
associated with the regularisation of both \sic\ and non--\sic\
theories.  These issues are explained in Sec.~2.

Even with the universal form for the soft breakings alluded to above,
there is still a lot of parameter--space. In Sec.~3 it
is explained that with the further assumption that
in the underlying theory the universal form of the soft terms is
invariant under renormalisation, the sparticle spectrum becomes
entirely determined by a single parameter. This {\it strong\/}
universality might be a property of the fundamental theory, or
it might arise to a good approximation
in the infra--red limit at $M_G$, from a more general
class of theories at higher scales. The results for
the SSM are explored in Sec.~4.

\section{DRED (Scylla) and DREG (Charybdis)}
\setcounter{equation}{0}
\renewcommand{\theequation}{2.\arabic{equation}}
Dimensional regularisation (DREG) is inconvenient for supersymmetric
theories. The fact that, for example, the quark--quark--gluon and
the quark--squark--gluino couplings are equal (because of
\sy) is not preserved under
 renormalisation, if DREG is employed.
If we demand that the two renormalised couplings are
the same, then the associated subtractions are different: or, to put it
another way, if the couplings are equal at one renormalisation scale,
$\mu$,   then they are different at another. This point is academic if
we  are calculating at a single value of $\mu$, but becomes  important
if we want to relate  a given theory at one value of $\mu$ to the same
theory at another such value:  as when we perform the standard running
analysis.  What this means is that DREG is    very inconvenient for the
SSM. If we assume  ``convenient'' values for the couplings at
unification (such as equality for the  couplings mentioned above) then
these couplings will be different at $M_Z$ and this difference will have
to be accounted for both in the actual
evolution analysis and in the calculation of the physical masses.

In 1979 Siegel\cite{siegel} proposed a modification of DREG designed to
render it more  compatible with \sy. The essential difference between
Siegel's  method (DRED\footnote{DRED is a sympathetic antipathy
and an antipathetic sympathy (Kierkegaard)} )
and DREG is that the continuation from $4$ to
$d$ dimensions  is made by {\it compactification\/} or {\it dimensional
reduction}.  Thus  while the momentum (or space-time) integrals are
$d$-dimensional in the  usual way, the number of field components
remains unchanged and consequently  supersymmetry is undisturbed.

 Modulo certain ambiguities that do not manifest themselves at  ordinary
loop levels, DRED is a practical \sic\ regulator. So  practical, in fact
that it has sometimes been used as being simpler  than DREG even
for  non--\sic\ theories  such as QCD. That DRED is a viable alternative
to DREG has long been believed\cite{capper}; but there are
subtleties involved  that have only been resolved
recently\cite{jjra,jjrb}. These arise due to the  effect of
Siegel's compactification on the gauge fields. After dimensional
reduction to $d = 4 - \epsilon$, it is only the first $d$ components of
the gauge field $A_{\mu}(x)$ that form the actual gauge connection. The
remaining $\epsilon$ components transform under gauge transformations
 as a multiplet of scalar fields, called $\epsilon$-scalars.

Now in a straightforward implementation of DRED in, for example, QCD,
the quark--quark--gluon and the quark--quark-$\epsilon$-scalar  coupling
are both equal to the gauge coupling. This equality is not preserved
under  renormalisation, however, because the latter interaction
 is independently gauge invariant.  We call interactions involving the
$\epsilon$-scalars {\it evanescent\/} interactions. Only in a
\sic\ theory do they remain equal to their ``natural'' values
under renormalisation. If we denote the
genuine masses and couplings  of a theory collectively as $\la$ and
the evanescent ones as $\la_E$, then it is possible to show that
the $S$-matrix (${\cal S}$) is independent
of $\la_E$ in the sense that there exists
 a coupling constant redefinition
\be
\lambda' = \lambda'(\lambda, \lambda_E) \quad\hbox{and}\quad
\lambda'_E = \lambda'_E (\lambda, \lambda_E)
\ee
such that we have
\be
{\cal S}(\la) = {\cal S}_{\rm DRED}(\la', \la'_E)
\ee

This had to be the case, of course, for DRED to be a consistent
regulator. Evidently varying $\la_E$ defines a trajectory in $(\la',
\la'_E)$-space without changing the $S$-matrix. It follows that we are
free to choose a point on this trajectory such that the $\la'_E$ are
indeed equal to their natural values. If this is done, however, it
should be clear that it would {\it not\/} be possible (using DRED) to
relate predictions made at different   values of the renormalisation
scale $\mu$ by evolving only the $\beta$-functions corresponding to the
real interactions.

To sum up: DREG is inconvenient for a running analysis in a \sic\ theory
because coupling constant relations prescribed by \sy\ are not
preserved, while DRED is inconvenient for non--\sic\ theories because
evanescent couplings do not remain equal to their natural values,
and enter into the $\beta$--functions for the genuine couplings.
This seems to leave us with an obvious  choice for any given theory;
but, as we shall see in the next section, the case of the
SSM presents special problems.

\section{The \ssm}
\setcounter{equation}{0}
\renewcommand{\theequation}{3.\arabic{equation}}

Let us consider the standard running analysis from $M_G$ to $M_Z$ in the
SSM, starting with the dimensionless couplings. If we use the  whole SSM
as our effective field theory throughout, then there is no   need to
introduce evanescent dimensionless couplings, because as  far as the
dimensionless coupling sector is concerned the theory is  effectively
\sic .   We can with confidence proceed to include two--loop
contributions to the $\beta$--functions. One must ensure  that
the input values of the couplings at $M_Z$ are those appropriate  to the
SSM rather than the standard model, which means they will depend
through  radiative  corrections on the sparticle spectrum\cite{bmp}.

There is an alternative approach whereby for scales below any given
particle mass, $M_S$ say,  the contribution  for the corresponding
particle is excised from  the $\beta$--functions; in other words, below
each particle mass a new effective theory is defined with the said
particle integrated out.   Evidently this  approach sums to all orders
contributions of the form $\alpha\ln (M_S / M_Z)$ but neglects
non--logarithmic terms that are equally important unless  $M_S >> M_Z$.
Within the context of the effective field theory approach  it is
difficult to recover these non--logarithmic terms; one need only reflect
that the true effective theory below $M_S$ contains  nonrenormalisable
interactions which are suppressed only by powers of  $M_Z / M_S$.
Another criticism of this approach is that once we start decoupling
particles we lose \sy\ and thus to go beyond one loop we would need  to
address the evanescent coupling problem explained in the previous
section.  It therefore appears preferable to work throughout with the
SSM as the  effective field theory.

In fact, of course, the SSM is not fully \sic\ because of the soft
breaking terms, and so when we come to run the masses we cannot avoid
worrying about the $\epsilon$-scalars\cite{jja}. The reason is that
since they are    indeed scalars, there is no symmetry which forbids
them from having a mass.  If we set this mass zero at (say) $M_G$ then it
will be non--zero at $M_Z$,  and it will also influence (at two--loops)
the evolution of the genuine  scalar masses.  This is not a problem
 in principle, but it is more convenient to make a slight change in the
regularisation scheme\cite{jjmmy} which decouples the $\epsilon$-scalar
masses from the $\beta$--functions for the genuine scalar masses. The
same redefinition renders the one--loop pole masses for the scalars
independent of the $\epsilon$-scalar mass.

One might wonder whether it might not be simpler to employ DREG since
then the $\epsilon$-scalars do not appear at all. The problem then,
however,  is that the evolution of the dimensionless couplings would
become more  complicated, as explained at the beginning of the last
section. In subsequent  sections we implicitly assume use of the hybrid
scheme\cite{jjmmy} as  indicated above.

\section{Universality}
\setcounter{equation}{0}
\renewcommand{\theequation}{4.\arabic{equation}}

In this section we describe  how a particular ``universal'' form for the
soft-breaking couplings in a softly broken $N=1$
theory is renormalisation-group invariant through two loops, provided we
impose one simple condition on the dimensionless couplings\cite{jjb}.
The universal form for the trilinear couplings and $\phi^*\phi$  mass
terms is identical to that found in a derivation of the soft-breaking
terms from string  theory\cite{ilkiblm}.

We begin with a general $N=1$ \sic\ gauge theory.
The Lagrangian $L_{\rm SUSY} (W)$ is defined by the superpotential
\be
W={1\over6}Y^{ijk}\Phi_i\Phi_j\Phi_k+{1\over2}\mu^{ij}\Phi_i\Phi_j.
\ee
$L_{\rm SUSY}$ is the Lagrangian for  the $N=1$ supersymmetric gauge
theory, containing the gauge multiplet $\{A_{\mu},\la\}$ ($\la$ being
the gaugino) and a chiral superfield $\Phi_i$ with component fields
$\{\phi_i,\psi_i\}$ transforming as a (in general reducible)
representation $R$ of the gauge group $\cal G$. We assume that there are
no gauge-singlet fields and that $\cal G$ is simple. (The generalisation
to a semi-simple group is trivial.)  The soft breaking is incorporated
in $L_{\rm SB}$, given by

\be
L_{\rm SB}=(m^2)^j_i\phi^{i}\phi_j+
\left({1\over6}h^{ijk}\phi_i\phi_j\phi_k+{1\over2}b^{ij}\phi_i\phi_j
+ {1\over2}M\la\la+{\rm h.c.}\right)
\ee
(Here and elsewhere, quantities with superscripts are complex conjugates of
those with subscripts; thus $\phi^i\equiv(\phi_i)^*$.)

Given a certain constraint on the dimensionless couplings,
the following relations among the soft breakings
are renormalisation group invariant through two--loops\cite{jja}:
\bea
h^{ijk} & = & -MY^{ijk},\label{eq.hma}\\
(m^2)^i{}_j & = &
{1\over3}(1-{1\over{\lf}}{2\over3}g^2Q)MM^*\delta^i{}_j \\
b^{ij} & = & -{2\over3}M\mu^{ij}.
\label{eq.hmb}
\eea
The aforementioned constraint is
\be
P^i{}_j={1\over3}g^2Q\delta^i{}_j,\label{eq.pq}
\ee
where
\be
Q =  T(R)-3C(G),\quad\hbox{and}\quad
P^i{}_j =  {1\over2}Y^{ikl}Y_{jkl}-2g^2C(R)^i{}_j.
\ee
Here

\be
T(R)\delta_{AB} = \Tr(R_A R_B),\quad C(G)\delta_{AB} =
f_{ACD}f_{BCD} \quad\hbox{and}\quad
C(R)^i{}_j = (R_A R_A)^i{}_j,
\ee
where the $f_{ABC}$ are the structure constants of ${\cal G}$.

In the usual SSM notation, Eqs.~(\ref{eq.hma})-(\ref{eq.hmb})\ correspond to
a universal scalar mass $m_0$  and universal $A$ and $B$ parameters
related (to lowest order in $g^2$) to the gaugino mass $M$ as follows:
\bea
 m_0  & = & {1\over{\sqrt{3}}}M,\label{eq.maba}\\
A & = & -M,\label{eq.mabb}\\
B & = & -{2\over3}M.\label{eq.mabc}
\eea
Remarkably,  relations of this form can arise in effective supergravity
 theories motivated by superstring theory, where \sy\ breaking is
assumed to occur purely via vacuum expectation values for dilaton  and
moduli  fields\cite{ilkiblm}. Eqs.~(\ref{eq.maba}) and (\ref{eq.mabb})
are of fairly general  validity in this context; the relationship
between $B$ and $M$ is more  model dependent. Given certain assumptions
including  dilaton dominance  the result is $B = 2M/\sqrt{3}$; this case
 has been subject to some phenomenological investigation\cite{blm}.
The similarity between the conditions on the soft-breaking terms
which arise from our universality hypothesis and those that emerge
from string theory is certainly intriguing.
Eqs.~(\ref{eq.maba}) and (\ref{eq.mabb}) also arise\cite{jmy}
 in the context of {\it finite\/} \sic\ theories (which correspond to
a special case of Eq.~(\ref{eq.pq}), $P=Q=0$). (Recently Ib\'a\~nez has
discussed whether emergence of a finite low energy
effective field theory  from a string theory might be natural\cite{ibanez}.)

There is, however, an alternative interpretation of the above results.
Consider a unified theory where it would be {\it possible\/} to impose
Eq.~(\ref{eq.pq}). The fact that Eqs.~(\ref{eq.pq}) and
(\ref{eq.maba})--(\ref{eq.mabc}) are renormalisation group invariant is of
course equivalent to saying that they are  fixed points of the evolution
equations; fixed points, moreover, that are approached, given certain
conditions, in the
infra--red. For example, given a theory based on a simple group
with a single Yukawa coupling and a chiral multiplet transforming as
 an irreducible representation $R$, then Eqs.~(\ref{eq.pq}),
(\ref{eq.maba}) and (\ref{eq.mabb}) are infra--red attractive
as long as $-6C(R) < Q < 6C(R)$,  while (\ref{eq.mabc}) is too if $Q < 0$.
At first sight it might appear that the difference between
$M_P$  and $M_G$ is insufficient to allow significant evolution, but it
has recently  been argued\cite{ross} that in the case of the Yukawa
couplings the evolution towards the fixed point may occur more rapidly  in
the unified theory than in the low energy theory. If we believe that
this  conclusion holds also for the soft terms, then it is possible to
argue that for a wide range of  input parameters the boundary conditions
(\ref{eq.maba})--(\ref{eq.mabc}) might hold at $M_G$. (Since, however, $Q > 0$
is favoured for rapid evolution\cite{ross} we may have problems with
Eq.~(\ref{eq.mabc}).)

Let us turn now to phenomenology\cite{jjrc}.  We assume that the SSM is
valid below gauge unification, and that the unified theory satisfies
Eq.~(\ref{eq.pq}), either exactly or in the infra--red limit at $M_G$.
We then proceed to impose
Eqs.~(\ref{eq.maba})-(\ref{eq.mabc})  as boundary conditions at the
gauge unification scale.

\section{The running analysis}
\setcounter{equation}{0}
\renewcommand{\theequation}{5.\arabic{equation}}

We start with the superpotential:
\be
W = \mu_s H_1 H_2 + \lambda_t H_2 Q\tbar  + \lambda_b H_1 Q\bbar
+ \lambda_{\tau} H_1 L\taub\label{eq.susyw}
\ee
where we neglect Yukawa couplings except for those of the third
generation.
The Lagrangian for the SSM is  defined by the superpotential of
Eq.~(\ref{eq.susyw}) augmented with  soft breaking terms as follows:
\be
L_{\rm SSM} = L_{\rm SUSY}(W) + L_{\rm SOFT}
\ee
where
\bea
L_{\rm SOFT} = & - & m_1^2 H_1^{\dag}H_1 - m_2^2 H_2^{\dag}H_2
+ [m_3^2 H_1 H_2  + \hbox{h.c.} ]\nonumber\\
& - & \sum_i \left( m_Q^2 |Q|^2 + m_L^2 |L|^2
+m^2_{\tbar} |\tbar |^2 +m^2_{\bbar} |\bbar |^2
+ m^2_{\taub} |\taub |^2\right)\nonumber\\
& + & [ A_t\la_t H_2 Q\tbar + A_b \la_b H_1 Q\bbar +
A_{\tau}\la_{\tau} H_1 L\taub  +  \hbox{h.c.} ]\nonumber\\
& - & \half [ M_1\la_1\la_1 + M_2\la_2\la_2 + M_3\la_3\la_3
+  \hbox{h.c.} ]
\eea
and the sum over $i$ for the $m^2$ terms is a sum over the three generations.
The running analysis of the SSM has been performed
many times. The novel feature here is the restricted set of boundary
conditions at gauge unification, where we impose (in the usual notation)
\be
m_1 = m_2 = m_Q = m_L = m_{\tbar} = m_{\bbar}
= m_{\taub} = {1\over{\sqrt{3}}}M,\label{eq.mama}
\ee
\be
A_{\tau} = A_b = A_t = -M, \quad M_1 = M_2 = M_3 = M,
\ee
\be
m_3^2 = -{2\over 3}\mu_s M
\ee
where Eq.~(\ref{eq.mama}) includes the squarks and sleptons of all
three generations.

\begin{figure}
\center
\centerline{\psfig{figure=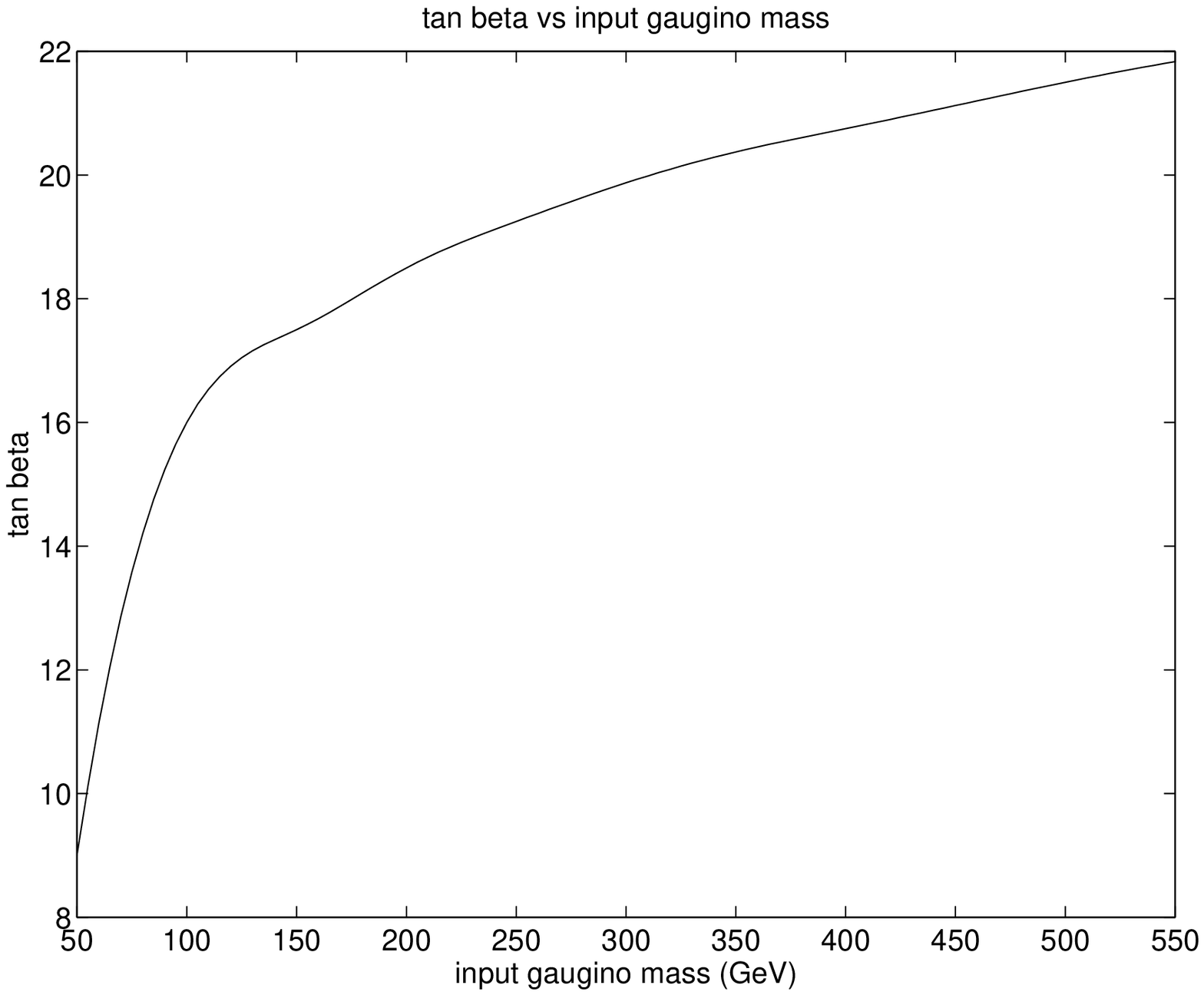,height= 4.0in}}
\fcaption{$\tan\beta$ vs. $M$ for  $m_t = 175$~GeV.}
\label{fig:tanbeta}
\end{figure}
Our procedure is as follows. We input $\alpha_1$, $\alpha_2$,
$\alpha_3$,  $m_t$ and $\tan\beta$ at $M_Z$, and  calculate the
unification scale  $M_G$ (defined as the meeting point of  $\alpha_1$
and $\alpha_2$) by running the dimensionless couplings. Then we input
the gaugino mass  $M$ at $M_G$, and run the dimensionful parameters
(apart from $m_3^2$ and $\mu_s$)  down to $M_Z$. We can then determine
$m_3^2$ and $\mu_s^2$ as usual  at $M_Z$ by minimising  the (one--loop
corrected) Higgs potential. Then we run $m_3^2$ and $\mu_s$ back  up to
$M_G$ (for the two possibilities of $\hbox{ sign }\,\mu_s$)  and calculate
$B' = B/M = (m_3^2)/(M\mu_s)$. By plotting $B'$ against the input  value
of $\tan\beta$ we can then determine whether (for a given input $M$)
there  exists a value of $\tan\beta$ such that Eq.~(\ref{eq.mabc}) is
satisfied.   Given a set $M, \tan\beta$ satisfying our boundary
conditions we  can calculate the sparticle spectrum in the usual way and
plot the resulting masses against $M$. We have included
one--loop corrections in the minimisation of the Higgs potential,
and in the calculation of the
mass ($m_h$) of the lighter  CP--even Higgs  boson.
Our results for other masses are based
on the tree  mass matrices but again  with all running parameters
evaluated at the scale $M$.
Since  the two--loop corrections to the $\beta$--functions are now available
\cite{jja,mvy} we incorporate these. In general
their effect is very small, being most noticeable in the Higgs sector;
although the mass of the lightest Higgs is essentially
unchanged, the other Higgs masses are increased by up to $10\%$
by the two loop corrections.   Of course for precise predictions,
we should also include threshold corrections.

In Fig.~(\ref{fig:tanbeta}) we plot $\tan\beta$
against the input gaugino mass, $M$,
having satisfied Eq.~(\ref{eq.mabc}). We find that the results for the masses
of the various particles
exhibit linear behaviour for a wide range of input gaugino masses.
Rather than give more figures, we therefore summarise our
results in Table~1, which
gives a good approximation  (within a few GeV) for
$100\hbox{ GeV }< M < 500\hbox{ GeV}$.

The phenomenology of our results is fairly typical.  For $M\approx
150$~GeV, for example,  we have a  stable neutralino at 55GeV, a
$\tau$-slepton at 80GeV, and the light Higgs at $115$~GeV. Notice that
$m_h$ is almost independent of $M$.   The main distinguishing feature of
our scenario lies in the relationship  between $\tan\beta$ and $M$, as
shown in Fig.~(\ref{fig:tanbeta}). At first  sight this appears to disfavour
$b-\tau$ unification. This is of course in any  case sensitive to the
nature of the unified theory which according to our scenario  is
required to satisfy Eq.~(\ref{eq.pq}).

\begin{table}[h]
\tcaption{Linear approximations of the form  $m = aM +b$
to the mass spectrum for
  $m_t = 175$~GeV, $m_t = 185$~GeV and $m_t = 190$~GeV.}
\begin{tabular}{|c|c|c|c|c|c|c|}
\hline
$m_t$ & \multicolumn{2}{|c|}{175} & \multicolumn{2}{|c|}{185} &
\multicolumn{2}{|c|}{190}\\
\hline
$m = aM + b$ & $a$ & $b$ & $a$ & $b$ & $a$ & $b$\\ \hline

\quad $m_h$ &
\quad 0.048 \quad & \quad 109 \quad &
\quad 0.060 \quad & \quad 108 \quad &
\quad 0.070 \quad & \quad 106 \quad\\ \hline

\quad $m_H$ & \quad 1.613 \quad & \quad 15 \quad & \quad 1.800 \quad &
\quad 7 \quad &\quad 1.870 \quad & \quad 5 \quad \\ \hline

\quad $m_A$ & \quad 1.585 \quad & \quad  8 \quad & \quad 1.782 \quad &
\quad 4 \quad &\quad 1.855 \quad & \quad 2 \quad \\ \hline

\quad $m_{H^{\pm}}$ & \quad 1.555 \quad & \quad 25 \quad & \quad 1.755 \quad &
\quad 20 \quad &\quad 1.829 \quad & \quad 17 \quad \\ \hline

\quad $m_{\etilde_1}$ & \quad 0.872 \quad & \quad 12 \quad & \quad 0.873 \quad
&
\quad 12 \quad &\quad 0.874 \quad & \quad 11 \quad \\ \hline

\quad $m_{\etilde_2}$ & \quad 0.666 \quad & \quad 12 \quad & \quad 0.667 \quad
&
\quad 12 \quad &\quad 0.668 \quad & \quad 12 \quad \\ \hline

\quad $m_{\nutilde_e}$ & \quad 0.930 \quad & \quad -22 \quad & \quad 0.930
\quad &
\quad -21 \quad &\quad 0.930 \quad & \quad -21 \quad \\ \hline
 \quad $m_{\tautilde_1}$ & \quad 0.830 \quad & \quad 31
\quad & \quad 0.852 \quad &
\quad 22 \quad &\quad 0.861 \quad & \quad 18 \quad \\ \hline
 \quad $m_{\tautilde_2}$ & \quad 0.615 \quad & \quad -11 \quad &
\quad 0.644 \quad &
\quad 1 \quad &\quad 0.657 \quad & \quad 5 \quad \\ \hline
 \quad $m_{\nutilde_{\tau}}$ & \quad 0.903 \quad &
\quad -21 \quad & \quad 0.917 \quad &
\quad -21 \quad &\quad 0.923 \quad & \quad -20 \quad \\ \hline
 \quad $m_{\chi_1^+}$ & \quad 1.527 \quad & \quad 48 \quad & \quad 1.580
\quad &
\quad 46 \quad &\quad 1.601 \quad & \quad 45 \quad \\ \hline
 \quad $m_{\chi_2^+}$ & \quad 0.793 \quad & \quad -21 \quad & \quad 0.799
\quad &
\quad -23 \quad &\quad 0.805 \quad & \quad -25 \quad \\ \hline
 \quad $m_{\chi_1^0}$ & \quad 1.532 \quad & \quad 44 \quad &
\quad 1.583 \quad &
\quad 44 \quad &\quad 1.603 \quad & \quad 45 \quad \\ \hline
 \quad $m_{\chi_2^0}$ & \quad 1.566 \quad & \quad 22 \quad & \quad 1.622
\quad &
\quad 20 \quad &\quad 1.645 \quad & \quad 18 \quad \\ \hline
 \quad $m_{\chi_3^0}$ & \quad 0.789 \quad & \quad -19 \quad & \quad 0.793
\quad &
\quad -20 \quad &\quad 0.797 \quad & \quad -21 \quad \\ \hline
 \quad $m_{\chi_4^0}$ & \quad 0.410 \quad & \quad -7 \quad & \quad 0.413
\quad &
\quad -8 \quad &\quad 0.417 \quad & \quad -9 \quad \\ \hline
 \quad $m_{\utilde_1}$ & \quad 2.264 \quad & \quad 26 \quad & \quad 2.266
\quad &
\quad 26 \quad &\quad 2.269 \quad & \quad 26 \quad \\ \hline
 \quad $m_{\utilde_2}$ & \quad 2.189 \quad & \quad 29 \quad & \quad 2.191
\quad &
\quad 30 \quad &\quad 2.194 \quad & \quad 30 \quad \\ \hline
 \quad $m_{\dtilde_1}$ & \quad 2.245 \quad & \quad 37 \quad & \quad 2.247
\quad &
\quad 37 \quad &\quad 2.251 \quad & \quad 37 \quad \\ \hline
 \quad $m_{\dtilde_2}$ & \quad 2.175 \quad & \quad 33 \quad & \quad 2.177
\quad &
\quad 33 \quad &\quad 2.180 \quad & \quad 33 \quad \\ \hline
 \quad $m_{\ttilde_1}$ & \quad 1.829 \quad & \quad 143 \quad & \quad 1.849
\quad &
\quad 143 \quad &\quad 1.861 \quad & \quad 142 \quad \\ \hline
 \quad $m_{\ttilde_2}$ & \quad 1.645 \quad & \quad 0 \quad & \quad 1.615
\quad &
\quad 18 \quad &\quad 1.609 \quad & \quad 27 \quad \\ \hline
 \quad $m_{\btilde_1}$ & \quad 2.040 \quad & \quad 56 \quad & \quad 2.113
\quad &
\quad 46 \quad &\quad 2.142 \quad & \quad 42 \quad \\ \hline
 \quad $m_{\btilde_2}$ & \quad 1.963 \quad & \quad 20 \quad & \quad 1.992
\quad &
\quad 28 \quad &\quad 2.009 \quad & \quad 30 \quad \\ \hline
\end{tabular}
\end{table}

\section{Outlook}

We have shown that the restrictions imposed by the conjecture of
renormalisation--invariant universality at  $M_G$ leaves a viable and
well determined \sic\ phenomenology. What we need now is a compelling
unified theory  that satisfies Eq.~(\ref{eq.pq}), (either exactly or  in
the infra--red).

\nonumsection{Acknowledgements}
I am grateful to Steve
Martin, Kevin Roberts, Mike Vaughn, Youichi Yamada and especially
Ian Jack for collaborations which led to the work described here. I
also thank Jon Bagger and the other PASCOS organisers for an
enjoyable meeting.

\nonumsection{References}

\end{document}